\documentclass[
]{revtex4-2}

\usepackage{acronym}
\usepackage{graphicx}
\usepackage{dcolumn}
\usepackage{bm}
\usepackage{orcidlink}
\usepackage{hyperref}
\usepackage{booktabs}
\usepackage{optidef}
\usepackage{dsfont}
\usepackage{xcolor}
\usepackage{placeins}
\usepackage[titletoc,toc,page,title,header]{appendix}
\usepackage{orcidlink}
\usepackage{ragged2e} 
\usepackage{braket} 
\usepackage{caption}
\usepackage{amsmath,amssymb, tikz}
\usetikzlibrary{quantikz2}

\mathtoolsset{showonlyrefs=true}

\newtheorem{definition}{Definition}

\hypersetup{
    colorlinks=true,
    linkcolor=blue,
    filecolor=magenta,      
    urlcolor=blue,
    citecolor=blue
}

\definecolor{bgblue}{RGB}{245,243,253}
\definecolor{ttblue}{RGB}{91,194,224}
\definecolor{deepblue}{rgb}{0,0,0.5}
\definecolor{deepred}{rgb}{0.6,0,0}
\definecolor{deepgreen}{rgb}{0,0.5,0}
\definecolor{deepgreen}{HTML}{395B64}
\definecolor{darkgreen}{HTML}{2C3333}
\definecolor{watermelon}{HTML}{FC6C85}
\definecolor{darkraspberry}{rgb}{0.53, 0.15, 0.34}

\newcommand{\PVI}{\mathrm{PVI}}

\newcommand{\I}{\mathcal{I}}

\newacro{PVI}{Pointwise $\mathcal{V}-$Information}
\newacro{QML}{Quantum Machine Learning}
\newacro{ML}{Machine Learning}

\begin{document}

\title{Which data samples matter when training a quantum model?}

\author{M. Emre Sahin\,\orcidlink{0000-0002-5996-0407}}
\affiliation{The Hartree Centre, STFC, Sci-Tech Daresbury, Warrington, WA4 4AD, United Kingdom}
\affiliation{Aston Institute of Photonic Technologies, Aston University, Birmingham, B4 7ET, UK}

\author{Cenk T\"uys\"uz\,\orcidlink{https://orcid.org/0000-0003-0257-9784}}
\affiliation{European Organisation for Nuclear Research (CERN), 1211 Geneva, Switzerland}

\author{David Galvin}
\affiliation{The Hartree Centre, STFC, Sci-Tech Daresbury, Warrington, WA4 4AD, United Kingdom}

\author{Oscar Wallis\,\orcidlink{0009-0002-7323-2059}}
\affiliation{The Hartree Centre, STFC, Sci-Tech Daresbury, Warrington, WA4 4AD, United Kingdom}

\author{Alex Gregory}
\affiliation{The Hartree Centre, STFC, Sci-Tech Daresbury, Warrington, WA4 4AD, United Kingdom}

\author{Paul Edwards}
\affiliation{The Hartree Centre, STFC, Sci-Tech Daresbury, Warrington, WA4 4AD, United Kingdom}

\author{Christa Zoufal,\orcidlink{0000-0003-4126-3141}}
\affiliation{IBM Research Europe – Zurich, Switzerland}

\date{\today}

\begin{abstract}
\textbf{The expressivity, loss landscapes, and generalization behavior of quantum machine learning models have been studied extensively, yet mainly from a model-centric point of view. Little is known about the compatibility of particular quantum models with specific datasets. In this work, we study quantum machine learning models with a data-dependent focus by establishing a quantum-compatible form of the $\mathcal{V}$-usable information. This metric quantifies the information a model can extract from a dataset, enabling a direct, data-centric comparison of different models. We employ this metric to study which data points contribute most significantly to a model's learning process. This enables us to develop an informed preselection strategy for data subsets that can significantly improve the data efficiency of the training pipeline. This work establishes the $\mathcal{V}$-usable information as a powerful diagnostic tool for learning dynamics and a practical route to data-efficient quantum machine learning.}
\end{abstract}

\maketitle


\begin{figure*}[t]
\centering
\includegraphics[width=\textwidth]{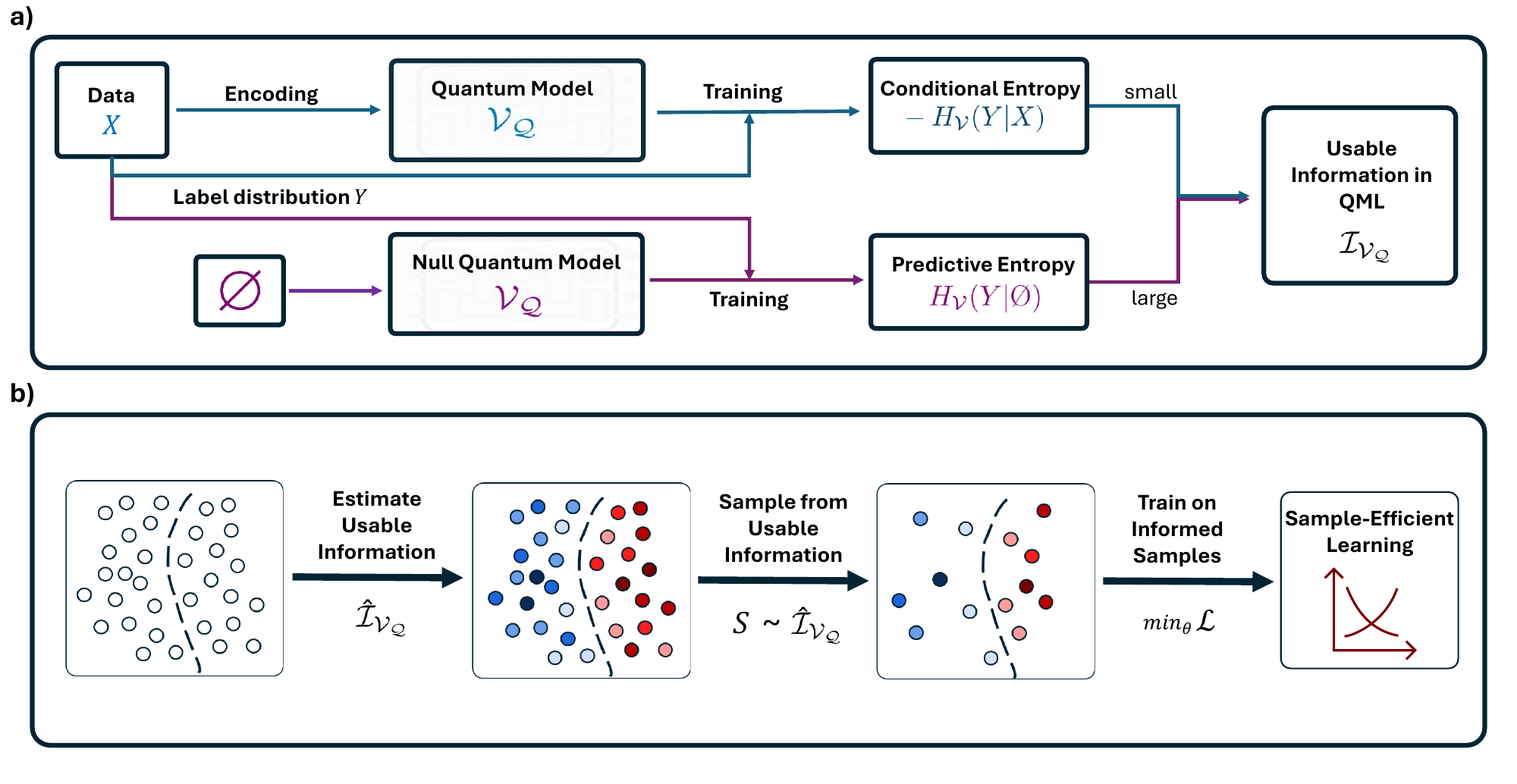}
\caption{\justifying Schematic of the $\mathcal{V}$‑usable information approach for quantum models, and its benefits. \textbf{a:} The dataset and an uninformative null-input are encoded into a quantum model $\mathcal{V}_Q$, respectively. After training, the conditional entropies can be estimated and used to evaluate the $\mathcal{V}$‑usable information. \textbf{b:} The metric can be used as a sampling method to have informative samples compared to uniform sampling. }

\label{fig:method-diagram}
\end{figure*}


\section{Introduction}
\label{sec:intro}

Quantum machine learning (QML) models have been argued to exhibit favourable representational properties and expressive power \cite{Vapnik2015,abbas2021power}.
Nevertheless, their practical advantages on real-world tasks remain unclear \cite{cerezo2022challenges}. Existing studies of QML have focused on the characterization of model expressivity, capacity, trainability, and generalization \cite{abbas2021power,du2022efficient,banchi2021generalization,huang2021power,caro2022generalization,caro2023information}. While these studies have brought important insights into the quantum models themselves, they do not directly address the compatibility of a particular model with a given dataset.
In this work, we investigate the expressivity of a QML model with respect to a given dataset and how an individual training sample contributes to the learning process. 

Our data-centric approach is based on the $\mathcal{V}$-usable information \cite{xu2019theory}, a metric that quantifies how much information about the labels is accessible to a restricted model family. 
This quantity measures information relative to the computational and architectural constraints of a chosen hypothesis class. 
It is a model-aware extension of Shannon's mutual information \cite{shannon1949mathematical,cover1999elements} with guarantees from the probably approximately correct (PAC) framework \cite{valiant1984learnable}, even in high-dimensional settings. 

We develop a $\mathcal{V}$-usable information implementation that extends the existing framework to quantum models. 
Our main contributions are the following. 
First, we formulate the $\mathcal{V}$-usable information for parametrized quantum models. The formulation is deliberately architecture-independent and therefore applies to feature-map--ansatz circuits, data re-uploading models, and integrated parametrized encodings alike. 
Second, we address the construction of a null input, which is an ingredient of the framework that does not carry over from the classical setting. 
Third, we study individual data samples using the pointwise $\mathcal{V}$-information (PVI)~\cite{ethayarajh2022understanding}, which assigns an information value to each training data point. This yields a fine-grained view of which examples are preferentially memorized, which support generalization, and how sample utility evolves during training.
Fourth, building on this sample-level view, we introduce a data-subset selection strategy that is informed by the PVI but does not require a trained model.
Finally, we demonstrate that both the diagnostic and selection pipeline remain tractable well beyond the reach of state-vector simulation by evaluating them for a $50$-qubit model using tensor-network methods. In this way, our work complements recent progress on model-dependent generalization theory by showing how pointwise, data-centric diagnostics can reveal structure that other metrics and generalization bounds may miss. 

The remainder of this work is structured as follows. Section~\ref{subsec:framework} introduces the $\mathcal{V}$-usable information framework and its extension to quantum models. Section~\ref{sec:results} presents our empirical study, beginning with the numerical methods (Sec.~\ref{sec:results:methods}). We then validate the quantum null input (Sec.~\ref{sec:null-input-results}), analyze learning dynamics (Sec.~\ref{sec:generalization-memorization}), and present our results on data selection (Sec.~\ref{sec:data-selection}). Finally, Section~\ref{sec:conclusion} summarizes our findings, discusses limitations, and outlines future work.

\section{\texorpdfstring{$\mathcal{V}$}{V}-usable Information for Quantum Machine Learning Models}
\label{subsec:framework}

The $\mathcal{V}$-usable information was introduced by \citet{xu2019theory} to quantify the information that a random variable $X$ provides for predicting another variable $Y$, relative to a predictive model family $\mathcal{V}$.

\begin{definition}[{\texorpdfstring{$\mathcal{V}$}{V}-Usable Information}]
Let $\mathcal{V}$ be a family of predictive models and $X$, $Y$ be random variables over sample spaces $\mathcal{X}$ and $\mathcal{Y}$, respectively. The \emph{$\mathcal{V}$-usable information} from $X$ to $Y$ is defined as
\begin{equation}
\mathcal{I}_{\mathcal{V}}(X \rightarrow Y) = H_{\mathcal{V}}(Y) - H_{\mathcal{V}}(Y|X),   
\end{equation}
where
\begin{equation}
H_{\mathcal{V}}(Y|X) = \inf_{f \in \mathcal{V}} \mathbb{E}\big[-\log f[X](Y)\big]
\end{equation}

is the conditional $\mathcal{V}$-entropy, and
\begin{equation}
H_{\mathcal{V}}(Y) = H_{\mathcal{V}}(Y|\emptyset)
= \inf_{f \in \mathcal{V}}
\mathbb{E}_{Y}\big[-\log f[\emptyset](Y)\big]
\end{equation}
is the predictive $\mathcal{V}$-entropy with $\emptyset$ denoting a null input that provides no information about $Y$.
\end{definition}

Intuitively, $\mathcal{I}_{\mathcal{V}}(X \rightarrow Y)$ measures the reduction in uncertainty about the labels $Y$ when the data $X$ is provided, but only the reduction that is achievable by a model within the family $\mathcal{V}$. The term $H_{\mathcal{V}}(Y)$ represents the model's minimum achievable prediction error (expected log loss) on $Y$ without any input (the prior), while $H_{\mathcal{V}}(Y|X)$ is the minimum error when the input $X$ is available. A large $\mathcal{I}_{\mathcal{V}}$ indicates high compatibility between the model family and the dataset's underlying structure.

The $\mathcal{V}$-usable information captures how $X$ informs us about $Y$ when subjected to computational constraints on transformations of $X$ imposed by $\mathcal{V}$. 
If $X$ does not provide any information about $Y$ in the context of the model class $\mathcal{V}$, then $\mathcal{I}_{\mathcal{V}}(X\rightarrow Y)=0.$
Otherwise, the $\mathcal{V}$-information will take any value in $\mathbb{R}^+$.
Unlike Shannon's mutual information, the $\mathcal{V}$-information is not symmetric and need not satisfy the data processing inequality.
In particular, preprocessing can increase the $\mathcal{V}$-usable information by making label-relevant structure more accessible to predictors in the chosen model family, even though it cannot increase the Shannon mutual information between the inputs and labels.

This dependence on the chosen model family also matters when comparing
$\mathcal{V}$-usable information values across models.
A larger value indicates a greater reduction in expected log loss
relative to the corresponding null-input baseline, so comparisons
across models require accounting for differences between their baselines.

The exact computation of the $\mathcal{V}$-usable information $I_{\mathcal{V}}(X \to Y)$ is generally intractable as it requires evaluating an infimum over the entire hypothesis class $\mathcal{V}$. In practice, we approximate this quantity empirically. Given a dataset $\mathcal{D} = \{(x_i, y_i)\}_{i=1}^{M}$ of size $M$, we estimate the predictive conditional entropy by training an ensemble of $K$ independent model instantiations, e.g., via gradient-based optimization from random parameter seeds. The empirical estimator is given by:

\begin{equation}
    \hat{\mathcal{I}}_{\mathcal{V}}(X \to Y) = H_{\mathcal{V}}(Y \mid \emptyset) + \frac{1}{M} \sum_{i=1}^{M} \log \left( \frac{1}{K} \sum_{k=1}^{K} P_{f_{\theta_k}}(y_i \mid x_i) \right),
\end{equation}

where $f_{\theta_k} \in \mathcal{V}$ denotes the quantum model parameterized by the $k$-th optimized or initialized parameter set $\theta_k$. 
The computational cost of this estimation scales as $\mathcal{O}(K \cdot M \cdot C)$, where $C$ represents the execution cost of the parameterized learning model per sample.

Furthermore, we can investigate the model's representational capabilities
for individual data points $x$ and corresponding labels $y$ using
pointwise $\mathcal{V}$-information (PVI), as introduced
by \citet{ethayarajh2022understanding}.
\begin{definition}[Pointwise \texorpdfstring{$\mathcal{V}$}{V}-Information]
Let $x$ be a data sample and $y$ a target class label. The pointwise \texorpdfstring{$\mathcal{V}$}{V}-Information (\emph{\ac{PVI}}) from $x$ to $y$ is defined as
\begin{equation}
    \I_\PVI(x \rightarrow y) = H_\mathcal{V}(y \mid \emptyset) - H_\mathcal{V}(y \mid x),
\end{equation}
where $H_\mathcal{V}(y \mid \emptyset)$ denotes the conditional $\mathcal{V}$ entropy of $y$ with no input, and $H_\mathcal{V}(y \mid x)$ denotes the conditional $\mathcal{V}$ entropy of $y$ conditioned on the input $x$.

\end{definition}

In practice, averaging the PVI over a subset of data points yields an empirical estimate of the $\mathcal{V}$-usable information. Contrary to the $\mathcal{V}$-usable information, the PVI can be negative, which suggests that the model predictions could be more accurate if certain data instances were neglected.
A negative PVI for a sample $(x,y)$ could be due to noisy, mislabeled, or particularly atypical data instances that conflict with the model's learned patterns, making them difficult to learn. 

Moreover, we can use the $\mathcal{V}-$usable information to investigate how well the model generalizes to unseen data. More specifically, we can investigate the absolute difference of the $\mathcal{V}-$usable information evaluated on train and test data, respectively. This enables us to compare the data-dependent model expressivity for the trained setting as well as on unseen data. 

\begin{definition}[\texorpdfstring{$\mathcal{V}$}{V}-Generalization Proxy]
\label{def:gen_proxy}
Let $\mathcal{V_Q}$ be a family of predictive models and $X$, $Y$ be random variables over $\mathcal{X} \times \mathcal{Y}$. The $\mathcal{V}$-generalization proxy is defined as
\begin{equation}
\Delta_{\mathcal{V_Q}} = \left| \mathcal{I}_{\mathcal{V_Q}}^{\mathrm{test}}(X \rightarrow Y) - \mathcal{I}_{\mathcal{V_Q}}^{\mathrm{train}}(X \rightarrow Y) \right|,
\label{eq:pvi-gap}
\end{equation}
where the superscripts indicate that each term is computed under the respective test or training distribution over $\mathcal{X} \times \mathcal{Y}$.
\end{definition}

To extend these definitions to quantum learning models, we consider
a classical data point $x \in \mathcal{X}$ and $k$ possible output labels.
We let $\mathcal{V_Q}$ denote a family of predictive parametrized quantum models. 
Each model produces an input-dependent quantum output $g_{\theta}(x)$, which a prediction rule $f$ maps to a probability distribution over labels.
Thus, $f_y(g_{\theta}(x))$ denotes the probability assigned to class $y$ given input $x$.
This notation is architecture-independent, since $g_{\theta}(x)$ may arise from a simple feature-map-ansatz circuit, a data re-uploading circuit, an integrated parametrized encoding, or another quantum learning architecture. 

The prediction rule $f : \mathbb{C}^{2^S} \rightarrow \mathbb{R}^k$
acts on a quantum state and returns a nonnegative, $\ell_1$-normalized
output vector whose $y$-th component
$f_y\left(\ket{g_{\theta}(x)}\right)$ encodes the probability
of the model assigning class $y$ to input $x$.
The corresponding conditional entropy can now be explicitly written as

\begin{equation}
\label{eq:cond_entr}
H_{\mathcal{V_Q}}(Y|X) = \inf_{f \in \mathcal{V_Q}} \mathbb{E}_{(x,y) \sim \mathcal{X} \times \mathcal{Y}}\left[ -\log f_y\left( \ket{g_{\theta}(x)} \right) \right].    
\end{equation}

Similarly, the predictive entropy for the quantum model is expressed as
\begin{equation}
\label{eq:pred_entr}
H_{\mathcal{V_Q}}(Y) = H_{\mathcal{V_Q}}(Y|\emptyset) = \inf_{f \in \mathcal{V_Q}} \mathbb{E}_{y \sim Y}\left[ -\log f_y\left( |g_{\theta}(\emptyset)\rangle \right) \right].
\end{equation}
We can then evaluate the \(\mathcal{V}\)-usable information for the quantum model family as
\begin{equation}
\label{eq:v-info}
\mathcal{I}_{\mathcal{V_Q}}(X\rightarrow Y)= H_{\mathcal{V_Q}}(Y)-H_{\mathcal{V_Q}}(Y|X).
\end{equation} 

\section{Results}
\label{sec:results}

This section presents an empirical study of the $\mathcal{V}$-usable information applied to QML models. First, we introduce the numerical methodology. Second, we validate the null-input model used to estimate the quantity in a QML context.  
Next, we test whether the \ac{PVI} resolves structural effects in model memorization and generalization at the level of individual samples. The respective experiments highlight the varying impact of different data samples from the same dataset. 
Finally, we investigate whether PVI can help reduce resource requirements
by improving the data efficiency of training.

\subsection{Methods}
\label{sec:results:methods}

\subsubsection{Datasets}
First, we consider the \textbf{phase-classification dataset} studied in Ref.~\cite{cong2019quantum,caro2022generalization,gil2024understanding}. The data correspond to ground states in different phases of a family of generalized cluster Hamiltonians given by
\begin{equation}
H = \sum_{i=1}^{S} Z_i
- J_1 \sum_{i=1}^{S-1} X_i X_{i+1}
- J_2 \sum_{i=2}^{S-1} X_{i-1} Z_i X_{i+1},
\end{equation}
where $Z_i$ and $X_i$ denote Pauli-Z and Pauli-X operators acting on the $i$-th qubit, respectively, and $S$ is the number of lattice sites arranged on a line. We consider a 1D system with open boundary conditions and 16 qubits. 
The coupling coefficients $J_1$ and $J_2$ determine the phase of the system. The phase diagram contains symmetry-protected topological, ferromagnetic, antiferromagnetic, and trivial phases~\cite{verresen2017one}. 

To generate the phase-classification dataset, we use the variational quantum eigensolver (VQE)~\cite{Peruzzo_2014} to estimate ground states of the Hamiltonian using the same hardware-efficient ansatz described above. The resulting VQE parameters are then used to prepare input states for the downstream QML models. In Qiskit, VQE optimization is carried out with the default constrained optimization by linear approximation (COBYLA) routine from SciPy~\cite{2020SciPy}. The training and test datasets consist of 240 and 1000 data points, respectively.

To ensure our findings are general, we also use two additional two-class classification benchmarks described as \textbf{adhoc} and \textbf{NTangled} datasets. 

The adhoc dataset offers a problem with synthetic ambiguity and redundancy. This is a two-dimensional classification task whose labels are induced by a quantum feature-map expectation landscape, following the conceptual framework of quantum-enhanced feature spaces~\cite{havlivcek2019supervised}. The task is defined by the expectation value
\[
s(\mathbf{x}) = \mathrm{Re}\langle \psi(\mathbf{x}) | \mathcal{O} | \psi(\mathbf{x}) \rangle
\]
of a random Hermitian observable $\mathcal{O}$ on the two-qubit state
\[
|\psi(\mathbf{x})\rangle =
U_{\Phi}(\mathbf{x}) H^{\otimes 2} U_{\Phi}(\mathbf{x})
|0\rangle^{\otimes 2},
\]
where $U_{\Phi}(\mathbf{x})$ denotes a diagonal $Z$-rotation and linear entanglement layer. Defining the margin as
$\mathcal{M}(\mathbf{x}) = |s(\mathbf{x})|$, we stratify the input space into three sub-populations based on margin quantiles: an \emph{easy} subset of high-margin points that are artificially duplicated to create redundancy, a \emph{hard} subset of rare but clean points that define the latent structure of the boundary, and a \emph{boundary} subset with stochastic label noise whose flip probability increases as the margin decreases. This construction provides a controlled setting in which redundancy, ambiguity, and sample difficulty can be varied independently.

The NTangled dataset~\cite{schatzki2021entangled} provides a quantum-specific classification problem on 8 qubits. The two classes are distinguished by Concentrated Entanglement (CE) values~\cite{beckey2021computable}: class 0 contains states with CE value 15 and class 1 contains states with CE value 45. The dataset contains 600 samples in total, split into 420 train and 180 test samples, and probes performance on highly entangled quantum states.

\subsubsection{Parameterized Quantum Model}
We consider the model class $\mathcal{V_Q}$ defined by a parameterized quantum circuit. 
For an input $x\in\mathcal{X}$, the circuit first prepares
\[
|\psi_x\rangle = U_x |0\rangle^{\otimes S},
\]
where $U_x$ is a data-dependent encoding unitary, and then applies a trainable unitary $\mathcal{G}_\theta$ to obtain
\[
|g_\theta(x)\rangle
=
\mathcal{G}_\theta U_x |0\rangle^{\otimes S}.
\]
The trainable unitary $\mathcal{G}_\theta$ consists of $10$ layers,
each applying a sequence of $R_Z$, $R_Y$, and $R_Z$ rotations
to every qubit, followed by CNOT gates arranged in a ring.
The CNOT gates are controlled by qubit $i$ and target qubit $i+1$
for $i=1,\ldots,S-1$, with an additional gate controlled by
qubit $S$ and targeting qubit $1$.
For example, when $S=16$, the three parametrized rotations per qubit in each
layer give $480$ trainable parameters.

The predictive distribution $p_\theta(y|x)$ is obtained from measurements of the final state. In the phase-classification setting, we are given input data which may be presented with $S=16$ qubits. The output should identify one of four labels. Hence, we measure the two central qubits $\frac{S}{2}=8$ and $\frac{S}{2}+1=9$. These two qubits can be measured in the following configurations $\set{00, 01, 10, 11}$, which in turn can be mapped to one of the four labels. We assign $p_\theta(y|x)$ as the number of times a particular label was sampled and divide it by the number of total samples.

All models were trained with a cross-entropy loss function 
\begin{align}
\mathcal{L}
= -\frac{1}{M}\sum_{i=1}^{M}\log p_\theta(y_i|x_i),
\end{align}
where $M$ denotes the number of training data samples, $y_i$ the target labels and $\hat{y}_i$ the model predictions.

Numerical simulations are carried out with Qiskit~\cite{qiskit}, Qiskit Machine Learning~\cite{2025-qiskit-ml-overview}, and quimb 1.12.1~\cite{gray2018quimb} and jax 0.10.1~\cite{jax2018github}. Parameter optimization is performed with the ADAM optimizer~\cite{kingma2014adam} using \texttt{amsgrad}~\cite{sashank2018convergence} and learning rate $0.1$.

\subsection{Quantum Null Input}
\label{sec:null-input-results}

Importantly, to evaluate the predictive entropy, we need to define a null input $\emptyset$ that provides no task-specific information about $Y$ while preserving the inductive bias of the model family. 
The corresponding null computation is denoted by $g_{\theta}(\emptyset)$, and its prediction by $f_y(g_{\theta}(\emptyset))$. 
More specifically, the aim of a null model is to construct an input to replace the data encoding unitary $U_x$ that does not provide any information about the data labels. 

In classical neural network applications setting all features to zero already cleanly provides a good null input as it isolates the model's bias term.
A natural fit for this replacement in a quantum model would be a depolarising channel. This would effectively replace the data input state with a maximally mixed state $I / 2^S,$ where $I$ is the identity matrix on the $2^S$ dimensional Hilbert space. 
In a unitary setting, the integration of this approach would require double the number of qubits, which is typically too significant an overhead for realistic implementations.
We aim for a feasible unitary adaptation $U_{\emptyset}$ and compare 3 possible candidates, as explained in Table \ref{tab:null-inputs}. We evaluate these candidate null encodings using the predictive entropy. 
The goal is to find an input that makes the model's output as close to a uniform--corresponding to maximum entropy--prediction as possible, thereby creating a non-informative baseline. A higher predictive entropy indicates a better null model.

\begin{figure}[!t]
  \centering
    \begingroup
    \centering
    \normalsize
    \textbf{Model}\par\vspace{3pt}
    \begin{quantikz}[column sep=0.45cm]
      \lstick{\makebox[1.4em][r]{$\lvert\psi\rangle$}}
      & \gate{\mathcal{G}_{\theta}}
      & \meter{}\arrow[r,"f"]
      & \push{\begin{pmatrix}P_1\\[-2pt]\vdots\\[-2pt]P_k\end{pmatrix}}\setwiretype{n}
    \end{quantikz}
    \par\vspace{12pt}
    \textbf{Null model}\par\vspace{3pt}
    \begin{quantikz}[column sep=0.45cm]
      \lstick{\makebox[1.4em][r]{$\lvert\psi_\emptyset\rangle$}}
      & \gate{\mathcal{G}_{\theta_\emptyset}}
      & \meter{}\arrow[r,"f"]
      & \push{\begin{pmatrix}P_1\\[-2pt]\vdots\\[-2pt]P_k\end{pmatrix}}\setwiretype{n}
    \end{quantikz}
    \par\vspace{8pt}
    \begin{quantikz}[column sep=0.4cm]
      \lstick{$\lvert\psi_\emptyset\rangle \equiv\lvert0\rangle^{\otimes S}$}
      & \gate{U_\emptyset} &
    \end{quantikz}
    \par
    \endgroup
  \caption{\textbf{Model and null model.} Both use the same parametrized quantum circuit architecture, measurement, and prediction rule $f$ to obtain an output probability vector. The null-input state is prepared as $\ket{\psi_\emptyset}=U_\emptyset\ket{0}^{\otimes S}$.}
  \label{fig:model-null-model}
\end{figure}
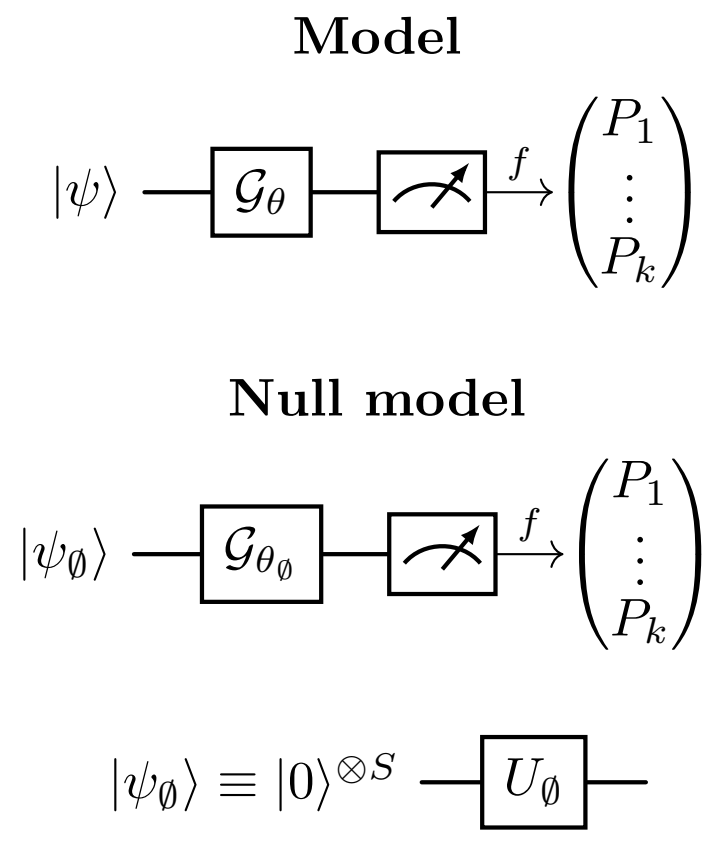

The null prediction $p_\theta(y|\emptyset)$ is obtained by replacing $U_x$ with three different null encoding $U_\emptyset$, while keeping the ansatz, measurement rule, and post-processing unchanged (Fig.~\ref{fig:model-null-model}).
For the first null encoding (N1), we prepare an approximate Haar random initialization. A true Haar-random null state on $n$ qubits is represented by the density matrix $\rho_{\emptyset} = \int_{\text{Haar}} |\psi\rangle \langle\psi| d\psi = \mathbb{I}/2^n$. We approximate this Haar-random baseline using a parameterized local random quantum circuit. Specifically, the state is prepared by applying a structured 5-layer circuit to $|0\rangle^{\otimes n}$, where each layer $l$ is partitioned into local single-qubit rotations and entangling blocks:

\begin{equation}
    U^{N1}_\emptyset = \mathcal{W}_{\text{alt}}^{(l)} \mathcal{R}^{(l, 2)} \mathcal{W}_{\text{nn}}^{(l)} \mathcal{R}^{(l, 1)}.
\end{equation}

For $\mu \in \{1, 2\}$, we apply independent, Haar-random single-qubit Euler rotations to all qubits:
    \begin{equation}
        \mathcal{R}^{(l, \mu)} = \bigotimes_{j=1}^{n} R_z(\alpha_{j, l, \mu}) R_y(\beta_{j, l, \mu}) R_z(\gamma_{j, l, \mu}),
    \end{equation}
    where the angles $\alpha, \beta, \gamma$ are sampled uniformly from $[0, 2\pi)$.
Furthermore, we apply 2-qubit controlled-NOT ($\text{CNOT}$) gates between adjacent qubits in a 1D linear chain topology:
    \begin{equation}
        \mathcal{W}_{\text{nn}}^{(l)} = \prod_{j=1}^{n-1} \text{CNOT}_{j, j+1}.
    \end{equation}
Finally, we aim to accelerate the scrambling and entanglement entropy growth across the system by applying $\text{CNOT}$ gates on alternating, offset pairs:
    \begin{equation}
        \mathcal{W}_{\text{alt}}^{(l)} = \prod_{j=1}^{\lfloor (n-1)/2 \rfloor} \text{CNOT}_{2j, 2j+1}.
    \end{equation}
This five-layer RQC structure forms a $t$-design ($t \geq 1$) that rapidly scrambles the input state, providing a pure-state surrogate that closely approximates the uniform projection of Haar-random states in the Hilbert space.
The second null encoding (N2) applies Hadamard gates on all qubits $U^{N2}_\emptyset = H^{\otimes n}$. The equal superposition state spreads amplitude uniformly across all basis states. When followed by entangling layers, it rapidly produces a predictive entropy close to the Haar-random limit without requiring purified ancilla qubits or density-matrix simulations.
Finally, we test a simple null encoding (N3) that does not apply any gates onto the zero-basis state $U^{N3}_\emptyset = \mathbb{I}$.

Based on the phase-classification dataset and the remaining methodology as described in Sec.~\ref{sec:results:methods} as an exploratory setting, we evaluate each null model on eight independently drawn random subsets, each containing 25\% of the training data. Table~\ref{tab:null-inputs} reports the mean pointwise predictive entropy on train and test samples. The Haar-random null (N1) and Hadamard null (N3) achieve similarly high predictive entropy, whereas the zero-state null (N2) performs substantially worse. Lower predictive entropy indicates that the null input retains residual structure that makes label prediction artificially easier and therefore provides a weaker baseline.
This result is not too surprising since setting the input to a fixed quantum state such as $|0\rangle^{\otimes n}$ or any other computational bases state can easily introduce a model bias through the ansatz layout $\mathcal{G}_\theta$. Instead, one would rather prioritize a null input that maximizes the predictive uncertainty under random parameterizations. The initialization through a Hadamard-product state $H^{\otimes n}|0\rangle^{\otimes n}$ (N3) effectively depolarizes single-qubit readout marginals across wide parameter regimes, mimicking the entropy of a maximally mixed state without requiring ancilla overhead.
Hence, we use N3 in the remainder of the paper. It achieves a similar values for $H_{\mathcal{V}}(Y)$ as N1 while being significantly easier to construct.

\begin{table}[t]
\centering
\begin{tabular}{llcc}
\toprule
Null input & Prepared state
& \multicolumn{2}{c}{$H_{\mathcal{V}}(Y)$} \\
\cmidrule(lr){3-4}
& & Training & Test \\
\midrule
N1 (Haar-random)
& $U_{\mathrm{Haar}}\ket{0}^{\otimes S}$
& $1.998 \pm 0.02$ & $1.998 \pm 0.02$ \\
\addlinespace
N2 (Zero state)
& $\ket{0}^{\otimes S}$
& $1.868 \pm 0.18$ & $1.873 \pm 0.15$ \\
\addlinespace
N3 (Hadamard)
& $H^{\otimes S}\ket{0}^{\otimes S}$
& $1.997 \pm 0.05$ & $1.997 \pm 0.05$ \\
\bottomrule
\end{tabular}
\caption{\textbf{Null-input states and predictive entropy.}
Mean pointwise predictive entropy on the training and test sets
of the phase-classification dataset for each candidate null input.
Here, $U_{\mathrm{Haar}}\sim\mathrm{Haar}\bigl(\mathrm{U}(2^S)\bigr)$.}
\label{tab:null-inputs}
\end{table}

\subsection{Understanding Learning and Generalization}
\label{sec:generalization-memorization}

\begin{figure*}[tbh!]
  \centering
  \includegraphics[width=0.9\linewidth]{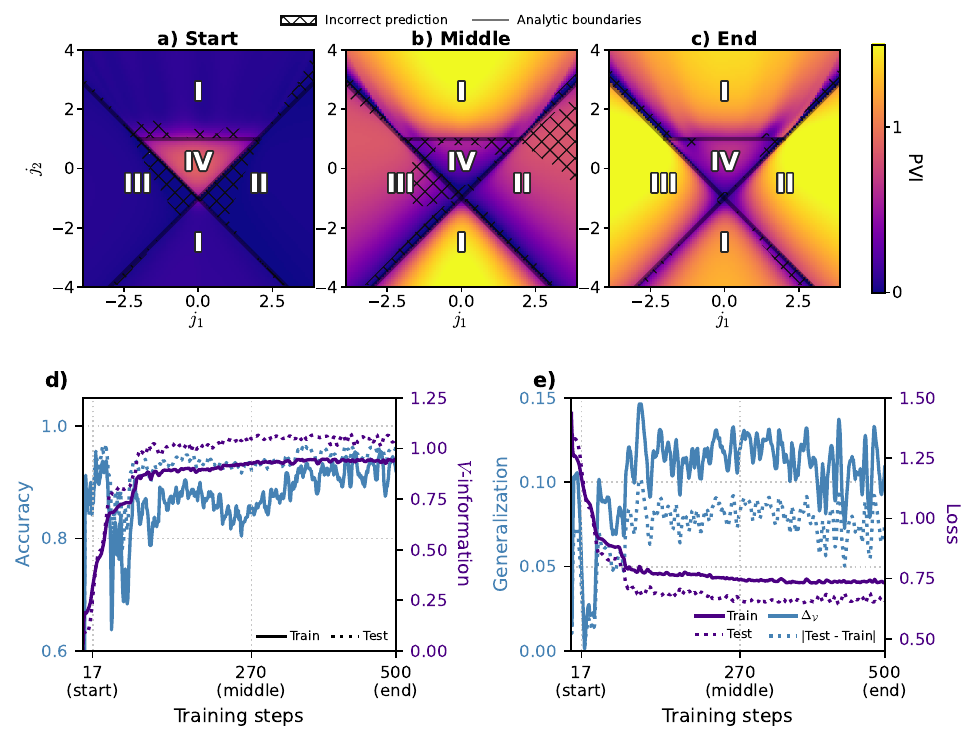}
  \caption{\justifying
  \textbf{PVI resolves sample-level learning structure in the phase-classification task.}
  \textbf{a--c}, PVI across the phase diagram at different training stages. \textbf{Grey lines indicate the theoretical phase boundaries from literature}, and crosses denote incorrectly classified points. Colours indicate PVI values, from low (blue) to high (yellow). \textbf{d}, Training and test accuracy (solid lines) alongside estimated $\mathcal{V}$-usable information (dotted lines). \textbf{e}, The $\mathcal{V}$-generalization proxy $\Delta_{\mathcal{V}}$ tracks the absolute difference in training and test loss.}
  \label{fig2}
\end{figure*}

Next, we test whether the $\mathcal{V}-$usable information resolves structural patterns in model learning and generalization that other metrics fail to capture. 
We train a parametrized quantum model on the phase-classification dataset--both as described in Sec.~\ref{sec:results:methods} using an Adam optimizer \cite{kingma2014adam}.

Fig.~\ref{fig2} a--c illustrates the PVI evaluated on the full dataset, i.e., training and test data, for the individual samples within the phase diagram. 
Lighter colors indicate higher PVI values for the respective data points and a better capability of the model to capture the underlying information.
At iteration $5$ of the training (start), the PVI remains low across the data space. Additionally, we can see that the model assigns the wrong labels to a significant fraction of samples within the full dataset.
After $50$ iterations (middle), the model is already able to capture phase I rather well.
When the training is completed, the PVI values reflect that the model was able to learn good representations for phases I to III.  However, the difficulty to learn the information of phase IV and the vicinity of phase boundaries is clearly represented by low PVI values. 
Additionally, we track the accuracy and the $\mathcal{V}$-usable information, approximated as the mean PVI across samples. 
Both metrics are evaluated throughout the training with respect to the training and the test dataset.
As shown in Fig.~\ref{fig2}d, train and test accuracy initially increase, then plateau, and finally fluctuate significantly. 
The  $\mathcal{V}$-usable information evaluated on train and test data, respectively, increases rather continuously and, thus, clearly reflects the effect of the training.

Finally, we compare the $\mathcal{V}$-generalization proxy defined in Def.~\ref{def:gen_proxy} with the difference in the cost function evaluated on train and test data, respectively.
As shown in Fig.~\ref{fig2}e, the $\mathcal{V}$-generalization proxy and the loss-based generalization proxy track one another qualitatively. The strong qualitative agreement between the two metrics provides empirical validation for $\Delta_{\mathcal{V}}$ as a meaningful, information-theoretic measure of a trained model's generalization performance. This suggests that the information the model can use from the test set, diverges from the training set in a manner proportional to the divergence in loss.

\begin{figure*}[tbh]
  \centering
  \includegraphics[width=0.9\linewidth]{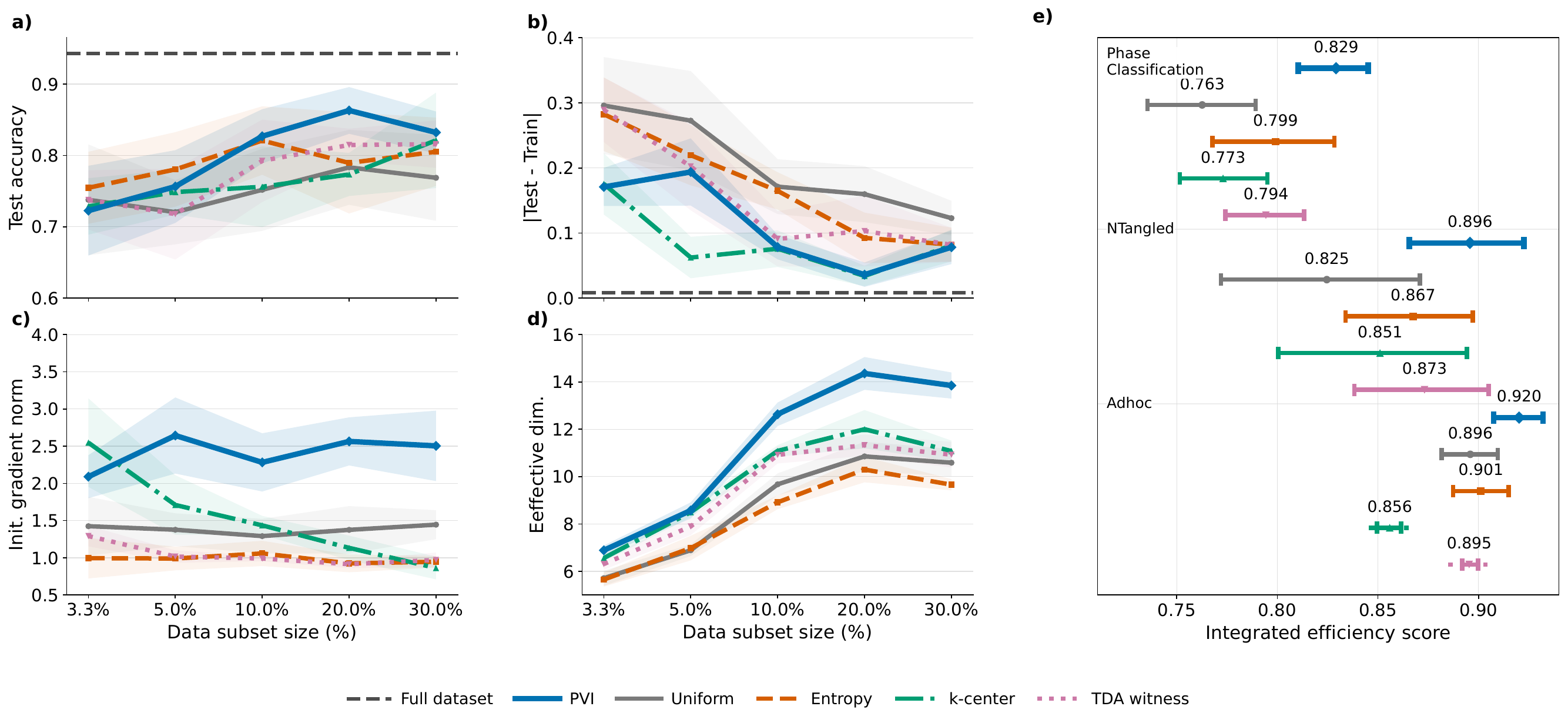}
  \caption{\justifying
  \textbf{$\mathcal{V}$-information-guided subset selection improves data efficiency in quantum machine learning.}
  Comparison of subset-selection strategies as a function of the fraction of the training set retained.
  \textbf{a)}, Accuracy evaluated on the test dataset. 
  \textbf{b)}, Absolute difference between the loss function evaluated on train and test datasets, respectively.
  \textbf{c)}, Gradient norm at initialization.
  \textbf{d)}, Effective dimension.
  \textbf{e)}, Cross-dataset summary of subset-selection efficiency across the phase-classification, adhoc \cite{havlivcek2019supervised}, and NTangled \cite{schatzki2021entangled} reported as a normalized integrated efficiency score.
 Dashed horizontal lines indicate the full-data reference where applicable. The curves represent the mean over 10 runs using randomly sampled parameter initializations, and the shaded regions denote the respective variance.}
  \label{fig3}
\end{figure*}

\subsection{Data Selection and Sampling}
\label{sec:data-selection}

In the following, we evaluate whether the PVI can guide the selection of a subset of training data that achieves comparable performance with reduced resources. In other words, we test if PVI-based selection can improve the data efficiency of a QML training pipeline.

The authors of Ref.~\cite{chen2025quality}, first train their model on the given dataset and, then, compute the PVI scores to identify samples with high information content. To select a reduced data subset, data samples with low or negative PVI scores are discarded. Their empirical results show that models trained on the identified data subset can match the performance of models trained on the full dataset. The main bottleneck of this method, however, is that it requires the training on the entire dataset which, in fact, is the resource intensive step that one wants to avoid effectively.

We introduce a PVI-informed data subset selection that circumvents this issue.
First, the PVI is computed for a subset of randomly selected parameter initializations without training the model.
Then, we construct a data subset based on the respective scores, where a data sample is more likely to be included if it corresponds to a lower PVI value. 
This choice is taken because low PVI values at initialization correspond to data samples that are not intrinsically well-represented by the inductive bias of the model. We hypothesize that focusing the training on these samples provides a stronger, more diverse learning signal, forcing the model to learn features it would not easily capture otherwise, ideally leading to better generalization.

Concretely, we compute the weight
\begin{equation}
w(x_i) \;=\; \max_{(x,y)\in\mathcal D}\,\widehat{\mathcal{I}}_{\rm PVI}(x,y)\;-\;\widehat{\mathcal{I}}_{\rm PVI}(x_i,y_i),
\end{equation}
for a data sample $x_i$ with label $y_i$ and $D$ representing the full dataset. We assign each instance a probability that determines its chances of being included in the data subset: 
\begin{equation}
\label{eq:samplingweight}
P(x_i \mid y_i) \;=\; \frac{w(x_i)}{\sum_{j} w(x_j)}.
\end{equation}

We benchmark the proposed \ac{PVI}-based subset-selection method for subsets of size $3.3\%$, $5\%$, $10\%$, $20\%$, $30\%$ w.r.t.~the full dataset against four established subset-selection strategies. 
\emph{Uniform random sampling}, \emph{Entropy sampling}~\cite{lewis1995sequential}, \emph{Core-set or $k$-centre sampling}~\cite{sener2017active}, and a \emph{Topological Data Analysis-inspired witness sampler}~\cite{de2004topological,sheehy2012linear}. 

\begin{figure*}[htb!]
  \centering
  \includegraphics[width=.95\linewidth]{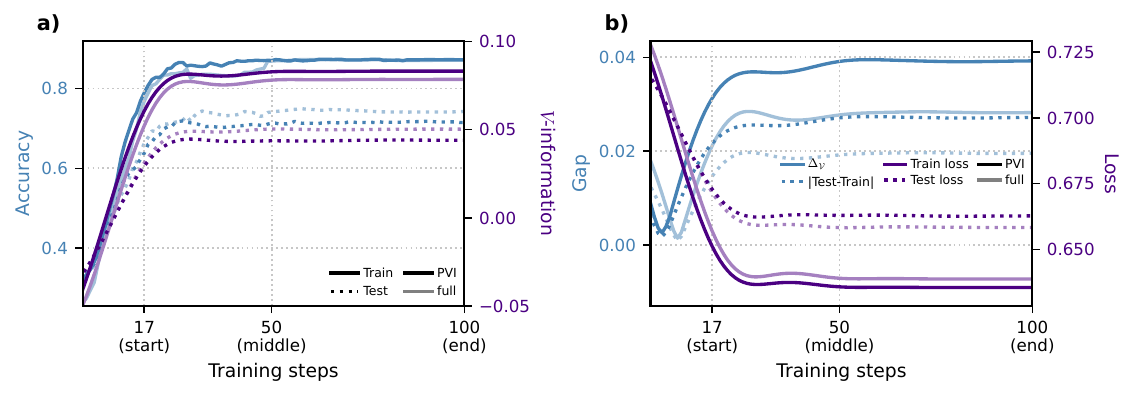}
  \caption{\justifying
  \textbf{Data-efficient training of 50-qubit model.} Model trained on $20\%$ of the dataset (darker colors) vs. model trained on full dataset (lighter colors)
\textbf{a)} Training and test accuracy together with the estimated $\mathcal{V}$-usable information during training. \textbf{b)} $\mathcal{V}$-generalization proxy and $\lvert \mathcal{L}_{\text{test}}-\mathcal{L}_{train}\rvert$ during training.}
  \label{fig4}
\end{figure*}

All methods are benchmarked against each other using the phase-classification dataset.
Across sampling budgets, selecting training points using our method yields strong results relative to baselines.
As shown in Fig.~\ref{fig3}, the PVI-guided data subset selection outperforms the other methods in the presented metrics if at least 10$\%$ of the dataset are selected.
As we can see from Fig.~\ref{fig3}c and d, the PVI-guided subset selection leads to comparatively larger gradients at model initialization and a higher effective dimension, which indicates favourable expressivity properties \cite{abbas2021power}.
To ensure that the observed behavior holds more generically, we also study the two additional datasets. 
The comparison of the sampling strategies across datasets and subset sizes with respect to the \textit{Integrated Efficiency Score} (IES) is presented in Fig.~\ref{fig3}e.
This metric summarizes predictive performance across training budgets by computing the normalized area under the test accuracy versus dataset size curve. For sampling
strategy $s$ and dataset $d$, we define

\begin{equation}
\label{eq:ies}
\mathrm{IES}_{s,d}
=
\frac{1}{b_m-b_1}
\sum_{i=1}^{m-1}
\frac{\bar{A}_i+\bar{A}_{i+1}}{2}
(b_{i+1}-b_i),
66\end{equation}

where $\bar{A}_i$ is the mean test accuracy at budget $b_i$
for strategy $s$ on dataset $d$. The score represents the
normalized area under the test-accuracy curve, evaluated
using trapezoidal integration. We report the bootstrap mean and 95\% percentile confidence interval. The normalization yields the average test accuracy over the evaluated budget range, allowing strategies to be compared without relying on a single subset size. Higher scores therefore indicate stronger predictive performance across the considered training budgets. Across all datasets, the PVI-guided data subset sampling consistently matches or exceeds competing strategies.

Now, we demonstrate a proof-of-concept example for the efficiency of PVI-based data subset selection at scale.
First, we generate $1000$ data points from the adhoc dataset per~\cite{havlivcek2019supervised} for 50 qubits. 
It should be noted that out of the three datasets studied in the PVI-analysis executed on smaller systems, this particular dataset is chosen for the scaled experiment as it can be generated trivially at various scales.  
Then, we use the methodology described above to select $20\%$ of the original data-points to train. Consistent with the method introduced above, the PVI scores used for selection are computed from a small set of randomly drawn parameter initializations, so no training on the full dataset is required at any point.

The parametrized quantum model differs from the remaining experiments in this work as it only employs a single layer. More specifically, the model applies the ZZ feature map of \cite{havlivcek2019supervised} followed by a single ansatz layer of parameterized $R_Y$ and $R_Z$ rotations and a nearest-neighbour $CZ$ chain, and reads out the total magnetization $z=\sum_i\langle Z_i\rangle$, which is mapped to a class probability by $p=\tfrac{1}{2}\bigl(1+\tanh(z/n)\bigr)$. The training uses ADAM with AMSGrad at learning rate $0.1$ for 500 steps, and all curves report the mean over 10 random parameter initializations.

Fig.~\ref{fig4} shows the resulting training trajectories for the model trained on the $20\%$ data subset in comparison to the model trained on the full dataset. All curves are means over ten independent runs. In Fig.~\ref{fig4}a the training and testing accuracy for the PVI-subset training reaches $0.874\pm0.024$ and $0.716\pm0.027$, respectively while the estimated $\mathcal{V}$-usable information rises from around zero at initialization to $0.083\pm0.004$ on the retained
training subset and $0.044\pm0.002$ bits on the test set. Training on the full dataset reaches a test accuracy of $0.742$. The difference of $2.6$ percent compared to the subset training is smaller than the standard deviation of $0.027$ across subset runs. 
Accuracy and
$\mathcal{V}$-information behave qualitatively similar and saturate early.
Fig.~\ref{fig4}b shows the $\mathcal{V}$-generalization proxy
$\Delta_{\mathcal{V}}$ for the subset training settles at $0.039\pm0.004$.
The illustrated loss curves are flat from roughly step $30$ onward for both the $20\%$ subset and full dataset training.
These results indicate that the PVI-based subset selection is indeed able to identify data samples which correspond to meaningful support vectors within the system.
Finally, it should be highlighted that the selected subset retains most of the full-dataset test accuracy while reducing the number of training samples by $80\%$.

\section{Conclusion and Discussion}
\label{sec:conclusion}

This work introduces the $\mathcal{V}$-usable information as a data-centric framework for analyzing quantum machine learning models. 
The presented extension of the $\mathcal{V}$-usable information framework to QML models, allows us to study effects in the training behavior that standard metrics such as accuracy could fail to capture. Furthermore, the metric provides an efficiently computable indication of a model's generalization behavior. 
Finally, the \ac{PVI} enables us to identify data samples that are particularly important for the model training and facilitates subset selection which can help to improve the data-efficiency of a learning process.

All in all, our results show that $\mathcal{V}$-usable information links information-theoretic ideas to practically measurable properties of quantum machine learning models. 
It provides a meaningful diagnostic of generalization behavior and yields a practical criterion for identifying training examples that matter most for performance. Although our experiments focused on quantum neural networks, the definitions and methodology extend naturally to other classes of QML models~\cite{amin2018, zoufal21, tuysuz2024, demidik2025, demidik2026}. Beyond the classification tasks considered here, a natural direction is to extend this framework to generative learning~\cite{tuysuz2026}, where training subset selection is also important for data efficiency.

Several limitations and open questions remain.
First, the metric is not normalized, and comparisons across tasks
or datasets require accounting for differences in label distributions
and null-input baselines.
Developing normalization methods could support broader comparative use.
Second, the null input is not uniquely defined for quantum models.
While our chosen construction is practical, this choice remains
a modelling assumption.
Further theoretical work could clarify how to choose null inputs
consistently across model families.
More broadly, future work should evaluate these metrics on other
QML models and assess their robustness to quantum hardware noise.

More generally, our findings reinforce the importance of empirical, data-dependent studies in QML. 
Theoretical guarantees do not always predict the practical behaviour of models trained on finite datasets.
In contrast, $\mathcal{V}$-usable information operates at the level of the learned predictor and the observed data distribution, enabling a finer-grained analysis of when and why quantum models generalize. 

We expect this perspective to help assess QML models and identify learning problems whose data structure is well matched to their predictive capabilities.

\begin{acknowledgments}
    We would like to thank Stefano Mensa, Elies Gil-Fuster and Stefan W\"orner for insightful feedback, and Edoardo Altamura for feedback and design of plots.
    This work was supported by the Hartree National Centre for Digital Innovation, a UK Government-funded collaboration between STFC and IBM. CT is supported by CERN through the CERN Quantum Technology Initiative. IBM, the IBM logo, and \href{https://www.ibm.com}{www.ibm.com} are trademarks of International Business Machines Corp., registered in many jurisdictions worldwide. Other product and service names might be trademarks of IBM or other companies. The current list of IBM trademarks is available at \href{https://www.ibm.com/legal/copyright-trademark}{www.ibm.com/legal/copytrade}.
\end{acknowledgments}

\bibliography{mainv3}

\end{document}